%Paper: hep-th/9307177
%From: Jim Horne <jhh@waldzell.physics.yale.edu>
%Date: Wed, 28 Jul 93 19:47:45 -0400

% This paper uses harvmac.tex and epsf.tex. There are two
% uuencoded tar-compressed figures that go with it. If you
% want to print out the figures properly with the paper,
% uncomment the following line.
%\input epsf

\input harvmac

\ifx\epsfbox\UnDeFiNeD\message{(NO epsf.tex, FIGURES WILL BE IGNORED)}
\def\figin#1{\vskip2in}% blank space instead
\else\message{(FIGURES WILL BE INCLUDED)}\def\figin#1{#1}\fi
\def\ifig#1#2#3{\xdef#1{fig.~\the\figno}
\goodbreak\midinsert\figin{\centerline{#3}}%
\smallskip\centerline{\vbox{\baselineskip12pt
\advance\hsize by -1truein\noindent\footnotefont{\bf Fig.~\the\figno:} #2}}
\bigskip\endinsert\global\advance\figno by1}

\def\ajou#1&#2(#3){\ \sl#1\bf#2\rm(19#3)}
\noblackbox
\def\a{\rightarrow}

\def\h{\hat}
\def\r{\hat r}

\def\t{\tilde}
\def\({\left (}
\def\){\right )}
\def\[{\left [}
\def\]{\right ]}
\def\RN{Reissner-Nordstr\"om}
\def\dtau{{\rm d}\tau^2}
\def\dr{{\rm d}r^2}
\def\drv{{\rm d}\vec{r}\cdot{\rm d}\vec{r}}
\def\dt{{\rm d}t^2}
\def\dht{{\rm d}\t t^2}
\def\dOmega{ r^2 {\rm d}\Omega^2}
\def\ds{{\rm d}s^2}
\def\dhs{{\rm d}\h s^2}
\def\conf{ -h \tau + 2 M/r}

%\font\ticp=cmcsc10
\gdef\journal#1, #2, #3, 19#4#5{
{\sl #1~}{\bf #2}, #3 (19#4#5)}

\lref\maki{T.~Maki and K.~Shiraishi, ``Extremal Black Holes and Strings
in Linear Dilaton Vacua,'' AJC-HEP-15, February 1993.}
\lref\hhs{J.H.~Horne, G.T~Horowitz, and A.R.~Steif,
\journal Phys. Rev. Lett., 68, 568, 1992, hep-th/9110065.}
\lref\kastor{D.~Kastor and J.~Traschen,
%``Cosmological Multi-Black Hole Solutions,''
\journal Phys. Rev., D47, 5370, 1993, hep-th/9112035.}
\lref\brill{D.R.~Brill, G.T.~Horowitz, D.~Kastor and J.~Traschen, ``Testing
Cosmic Censorship with Black Hole Collisions,'' NSF-ITP-93-78, UMHEP-387,
gr-qc/9307014.}
\lref\threed{G.T.~Horowitz and D. Welch,
%``String Theory Formulation of the Three-Dimensional Black Hole,
\journal Phys. Rev. Lett., 71, 328, 1993, hep-th/9302126;
N.~Kaloper, ``Miens of the Three-Dimensional Black Hole,'' Alberta-THY-8-1993,
hep-th/9303007.}
\lref\dark{G.T.~Horowitz, ``What is the True Description of Charged Black
Holes?,'' UCSBTH-92-52, gr-qc/9301008.}
\lref\maj{S.D.~Majumdar, \journal Phys. Rev., 72, 930, 1947;
A.~Papapetrou, \journal Proc. Roy. Irish Acad., A51, 191, 1947;
J. Hartle and S. Hawking, \journal Commun. Math. Phys., 26, 87, 1972.}
\lref\hay{D.R.~Brill and S.A.~Hayward, ``Global Structure of a Black
Hole Cosmos and its Extremes,''  gc-qc/9304007.}
\lref\pen{R.~Penrose,  {\it Rel. del Nuovo Cimento} {\bf 1}, 252 (1969).}
\lref\mye{R.~Myers, \journal Phys. Lett., 199, 371, 1987.}
\lref\aben{I.~Antoniadis,
C.~Bachas, J.~Ellis, and D.~Nanopoulos, \journal Phys. Lett., 211, 393, 1988;
\journal Nucl. Phys., B328, 117, 1989.}
\lref\sen{S.~Hassan and A.~Sen,
%``Twisting Classical Solutions in Heterotic String Theory,''
\journal Nucl. Phys., B375, 103, 1992, hep-th/9109038.}

\lref\moh{N.~Mohammedi, ``Naked Singularities in Four-dimensional
String Backgrounds,'' BONN-HEP-93-11, hep-th/9304013.}
\lref\witten{E.~Witten,
%``On String Theory and Black Holes,''
\journal Phys. Rev., D44, 314, 1991.}
\lref\ghs{D.~Garfinkle, G.~Horowitz and A.~Strominger,
%``Charged Black Holes in String Theory,''
\journal Phys.~Rev., D43, 3140, 1991; {\bf D45}, 3888({\bf E}) (1992).}
\lref\maeda{ G. Gibbons,
\journal Nucl. Phys., B207, 337, 1982;
G.~Gibbons and K.~Maeda,
\journal Nucl. Phys., B298, 741, 1988.}

\Title{\vbox{\baselineskip12pt\hbox{NSF-ITP-93-95}
\hbox{YCTP-P17-93}
\hbox{hep-th/9307177}
}}
{\vbox{\centerline {Cosmic Censorship and the Dilaton}
}}

\centerline{{James H. Horne}\footnote{$^\dagger$}
{Email address:
jhh@waldzell.physics.yale.edu}}
\vskip.1in
\centerline{\sl Department of Physics}
\centerline{\sl Yale University}
\centerline{\sl New Haven, CT 06511-8167}
\vskip .1in
\centerline{{Gary T. Horowitz}\footnote{$^*$}{On leave from the
Physics Department, University of California, Santa Barbara, CA
93106. Email address: gary@cosmic.physics.ucsb.edu} }

\vskip.1in
\centerline{\sl Institute for Theoretical Physics}
\centerline{\sl University of California}
\centerline{\sl Santa Barbara, CA 93106-4030}

\bigskip
\centerline{\bf Abstract}

We investigate extremal electrically charged black holes in
Einstein-Maxwell-dilaton theory with a cosmological constant inspired
by string theory. These solutions are not static, and a timelike
singularity eventually appears which is not surrounded by an event
horizon.  This suggests that cosmic censorship may be violated in this
theory.

\Date{7/93}
%\draft

\newsec{Introduction}

Perhaps the most important unsolved problem in classical general
relativity is cosmic censorship: Penrose's conjecture that generic
initial conditions do not evolve to form naked singularities~\pen. One
way to obtain some insight into this conjecture is to study exact
solutions.  Since cosmic censorship deals with generic initial data, a
particular solution, or even a family of solutions will never suffice
to prove or disprove this conjecture. However, they can be extremely
valuable in illustrating the type of behavior that is possible, and
may provide a starting point for a stability analysis that could
ultimately lead to a convincing counterexample.

Kastor and Traschen~\kastor\ have recently found an exact solution
describing colliding black holes. They consider charged black holes in
a universe with positive cosmological constant. This solution provides
a new arena to test cosmic censorship, and it has been shown that
naked singularities form in some of the black hole collisions~\brill.
In this paper we study a generalization of this solution which
includes a scalar dilaton coupled to the metric and Maxwell field in
the manner predicted by low energy string theory. There are several
motivations for doing so.  From the standpoint of general relativity,
the dilaton is a physically reasonable matter field, and one can
investigate whether cosmic censorship is valid for this choice of
matter.  From the standpoint of string theory, the dilaton is an
essential ingredient in the low energy theory. The analog of adding a
cosmological constant in string theory is to include excess central
charge. Solutions with excess central charge are of interest since
they arise naturally in several contexts, including gauged
Wess-Zumino-Witten models and noncritical string theories.  The
recently discussed exact black hole solutions in two~\witten\ and
three~\threed\ dimensions are examples of solutions with excess
central charge.

We will see that the dilaton changes the causal structure of the
Kastor-Traschen solution.  In particular, it turns out that in the
absence of any black holes, the solution with a dilaton describes a
collapsing Robertson-Walker universe with a final null singularity.
When even a single extremal\footnote{$^\dagger$}{We will show in
Sec.~2 that, unlike the Kastor-Traschen solution, the black holes in
this solution are extremal.} black hole is added, this null
singularity is replaced by a timelike one which begins far from the
black hole and is not hidden by an event horizon. (A more elaborate
model based on a generalized WZW construction seems to have similar
properties~\moh.) The solution with an
extremal black hole also has a null singularity at $r=0$, but we will
argue that this will probably be removed in the non-extremal solution.
Conformally rescaling to the string metric (the natural metric for
string theory) gives a spacetime which is asymptotically flat, and the
timelike singularity recedes to infinity.

The Kastor-Traschen solution~\kastor\ to the Einstein-Maxwell
equations with cosmological constant $\Lambda = 3h^2$ describes a
number of black holes, each having a charge equal to their mass,
situated at arbitrary locations.  The metric and gauge field for mass
parameters $M_i$ and positions $\vec{r}_i$ are given by
\eqn\kt{\ds =-{\dtau \over U^2} + U^2 \drv\,,
\qquad A_\tau ={1\over U} \,,}
where\footnote{$^\ddagger$}{Our expression
for $U$ differs from that of \brill\ by a factor of two multiplying $M_i$.
This is because we want $M =\sum M_i$ to represent the total mass of the
solution {\it with dilaton} (described below) in the limit when $h \a 0$.}
\eqn\udef{ U = -h\tau+\sum_i {2M_i\over |\vec{r}-\vec{r}_i|}\,.}
We will assume throughout that $h>0$.  The special case of a single
mass corresponds to the $Q=M$ \RN-de Sitter solution. The general case
is quite similar to the Majumdar-Papapetrou solution~\maj\ describing
a collection of $Q=M$ black holes in the absence of the cosmological
constant. If one simply sets $h=0$ in \udef, the resulting metric is
not asymptotically flat. To obtain the Majumdar-Papapetrou solution,
one must first introduce the coordinate $t$ measuring proper time at
infinity $ \tau = -e^{-ht}/h$, and then take the limit $h\rightarrow
0$ keeping $t$ fixed.  The resulting solution takes the form \kt\ with
$\tau$ replaced by $t$ and $U$ replaced by
\eqn\upm{ \t U = 1+\sum_i {2M_i\over |\vec{r}-\vec{r}_i|}\,.}

The low energy action that arises naturally in string theory with a
central charge $-h^2$ is
\eqn\saction{ S = \int {\rm d}^4 x \sqrt{-\h g} e^{-2\phi}[ -h^2 +
                      \h R + 4 (\nabla \phi)^2 - F^2 ]\,,}
where $\phi$ is the dilaton, $F$ is the Maxwell field, and $\h
g_{\mu\nu}$ is the string metric.  The $e^{-2 \phi}$ factor in front
of the scalar curvature can be eliminated by rescaling to the Einstein
metric using $g_{\mu\nu} = e^{-2 \phi} \h g_{\mu\nu}$ to yield
\eqn\eaction{ S = \int {\rm d}^4 x \sqrt{-g} [ R -2 (\nabla \phi)^2
   -h^2 e^{2\phi} - e^{-2\phi} F^2 ]\,.}
This can be viewed as general relativity with somewhat unusual matter.  The
constant $h$ provides a positive potential for the dilaton, so the
local energy density is always positive.  When $h=0$, the theory has a
duality symmetry which relates electrically charged solutions to
magnetically charged solutions and changes the sign of $\phi$. When
$h\ne 0 $, this symmetry is broken and there does not appear to be a
simple relation between solutions with different types of charge. We
will mainly discuss the electrically charged solution and comment on
the magnetically charged case at the end.

It was noticed in~\dark\ that an extremum of~\eaction\ with $h=0$ could
be obtained from the Majumdar-Papapetrou solution by setting
\eqn\dil{ e^{-2\phi} = \t U\,, \qquad A_t ={1\over \sqrt 2 \t U}\,,}
and taking the ``square root'' of the metric
\eqn\dmp{\ds   =-{\dt \over \t U} + \t U \drv\,.}
A single extremal charged black hole coupled to a dilaton has $Q^2= 2
M^2$.  The solution~\dmp\ describes a collection of $Q= \sqrt 2 M$
black holes which is static since the electromagnetic repulsion is
balanced by both the gravitational and dilatonic attraction. It can be
viewed as the dilatonic version of the Majumdar-Papapetrou solution.

Given the connection between these solutions when $h=0$, it is natural
to conjecture that the dilatonic version of the Kastor-Traschen
solution is
\eqn\dkt{\ds =-{\dtau \over U} + U \drv\,,
\qquad A_\tau ={1\over \sqrt 2 U}\,, \qquad e^{-2\phi} = U \,,}
where $U$ is again given by~\udef.  One can in fact verify that~\dkt\
is an extremum of~\eaction\ for any $h$. (This solution has been found
independently by Maki and Shiraishi~\maki, in different coordinates
that cover only part of the solution.)  Like the Kastor-Traschen
solution, \dkt\ is time dependent and has no symmetries in general.

In the next section we will explore the global properties of this
solution for a single mass. In section 3, we discuss the multi-mass
solution, as well as generalizations to higher dimensions and magnetic
charges.  Section 4 contains some concluding remarks.

\newsec{Single Mass Solution}

We begin by considering the solution~\dkt\ with no masses.  This will
clarify the type of universe in which the extremal black holes exist.
When $M_i = 0$, the metric~\dkt\ becomes
\eqn\flatein{ \ds =  {\dtau \over h \tau}
		-h \tau (\dr + \dOmega)\,,}
where $\tau <0$.  This can be put into a more familiar form by
defining a new coordinate $t =- \sqrt{-4 \tau/h}$.  The metric now
takes the form~\aben
\eqn\flateina{ \ds = - \dt
              + {1 \over 4} h^2 t^2 (\dr + \dOmega)\,.}
This is simply a collapsing Robertson-Walker universe with $k = 0$ and
$\rho = - 3 p$. The singularity at $t=0$ (which corresponds to $\tau =
0$) appears similar to the ``big crunch'' singularity of the standard
Friedmann solutions, but there is an important difference.  The
singularity in this solution is not spacelike but null! One way to see
this is to note that outgoing radial light rays satisfy $r= -{2\over
h}\ln |t| + r_0$, so that as $t \a 0, \ r\a \infty$. In other words,
observers never lose causal contact as they approach the
singularity. Another argument that the singularity must be null is
that the metric on the $r,t$ plane describes the interior of the past light
cone of the origin in two dimensional Minkowski spacetime, where $t$
labels the hyperbola. The surface $t=0$ is the light cone itself,
which is, of course, null. (If the spatial metric were not flat
Euclidean space, but rather the metric on a unit hyperboloid, the
spacetime metric would be flat everywhere.)  The Penrose diagram for
this spacetime is shown in Fig.~1.

\ifig\nomass{The Penrose diagram for the solution \dkt\ with no
masses. The dashed line at $r=0$ represents the origin of spherical
coordinates. The thick line at $\tau = 0$ is a null
singularity.}{\epsfbox{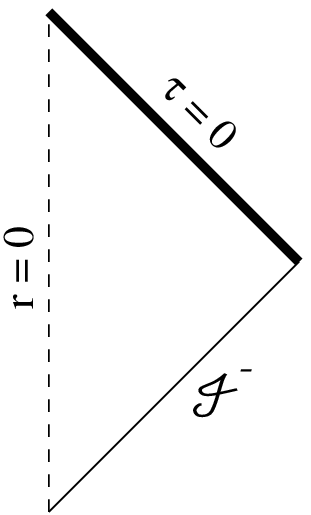}}

We now insert a single extremal electrically charged black hole into
this universe.  The metric is now given by
\eqn\einmetric{ \ds = - { \dtau \over \conf} +
                (\conf) ( \dr + \dOmega)\,.}
For $\tau < 0 $, the solution clearly approaches \flatein\ at large
$r$.  However as $\tau$ approaches zero, the presence of the mass has
an increasing effect even in the asymptotic region. Perhaps the most
important consequence of the mass is that the surface $\tau = 0$ is no
longer singular and no longer null. All constant $\tau$ surfaces are
spacelike, and observers can travel through $\tau = 0 $ without
difficulty.  However there is now a singularity along the curve $r
\tau = 2 M/h$.  We will see that this singularity has a completely
different character from the null singularity at $\tau = 0 $. In
particular, the new singularity is timelike and visible to observers
in the spacetime.

To show this, we consider the motion of radial null geodesics.  We
first verify that future directed null geodesics reach this
singularity in finite affine parameter, so the singularity is not ``at
infinity.'' We then show that past directed null geodesics can also
hit the singularity. This establishes that the singularity is timelike
and its effects can propagate into the spacetime.

Denote derivatives with respect to an affine parameter $\lambda$ by a
dot, $\dot{x}^{\mu} = { dx^{\mu} \over d\lambda}$, and consider only
null geodesics with $\dot{\theta} = \dot{\phi} = 0$. The fact that the
geodesic is null then requires
\eqn\nulldt{ \dot{\tau} = \pm (\conf) \dot{r}\,,}
where the plus (minus) sign is for future (past) directed outgoing
geodesics.  Since the metric is not static, there is no conserved
energy to use in calculating the geodesics. However a straightforward
calculation yields
\eqn\nullddr{ \ddot{r} + \Gamma^r_{\mu\nu} \dot{x}^\mu \dot{x}^\nu
	= \ddot{r} \mp h \dot{r}^2 = 0\,,}
where we have used~\nulldt. This equation can be easily solved to give
\eqn\eingeod{ r = \mp {1 \over h} \ln (c_1 \lambda + c_0)\,,}
where $c_0$ and $c_1$ are constants.  Since the singularity is at a
finite value of $r$, a future directed null geodesic reaches it in
finite affine parameter. Thus, the spacetime described by~\einmetric\
is geodesically incomplete.

Eq.~\eingeod\ also shows that past directed null geodesics reach
finite values of $r$ in finite affine parameter, but to show that
these curves hit the singularity, we need to explicitly find the
geodesics.  From \nulldt\ we see that past directed radial null curves
satisfy
\eqn\findge{ {d \tau\over dr} =  \(h\tau -{2M\over r}\)\,.}
This can be integrated to yield
\eqn\intgeo{ \tau = e^{h r}\( v - \int^r_{r_0}
     {2M\over \t r} e^{-h \t r} d\t r\)\,,}
where $v$ is a parameter labeling the different null curves, and
$r_0>0$ is an arbitrary constant. The integral on the right reaches a
finite limit as $r \a \infty$.  Denoting this limit $v_0$, we see that
the asymptotic behavior of the null curves depends crucially on
whether $v$ is greater than or less than $v_0$. For $v<v_0$, $\tau \a
-\infty$ as $r \a \infty$, and these geodesics reach past null
infinity. For $v>v_0$, $\tau$ clearly stays positive for all $r$ (and
in fact grows at large $r$). Since the singularity is at $r\tau =
2M/h$, these curves always hit the singularity. This shows that past
directed radial null geodesics can reach the singularity.

The fact that future directed null geodesics are incomplete is not too
surprising, since the $M = 0$ universe is also geodesically incomplete
due to the null singularity at $\tau = 0$.  However, the fact that
past directed null geodesics are also incomplete is quite unexpected.
This appears to be a serious violation of cosmic censorship. The
perfectly smooth universe in the far past~\flateina\ evolves into a
spacetime with a naked singularity.

Let us compare this behavior with other known solutions.  When there
is no dilaton, Kastor-Traschen solution \kt\ with a single mass still
has a timelike singularity at $r\tau = 2 M/h$. However, in that case
the singularity is either inside an event horizon (if $M$ is smaller
than an extremal value $M_{ext}$) or exists for all time (if $M >
M_{ext}$)~\hay.  So one does not have a violation of cosmic censorship
of the type found in the solution with a dilaton. In the \RN\
solution, one can consider a spacelike surface of constant $r$ with
$r_-< r < r_+$.  Since this surface is homogeneous, it can be viewed
as providing initial data for a homogeneous cosmology. If $Q=0$, the
universe collapses into a spacelike singularity at $r=0$ (the
Schwarzschild singularity).  However, if $Q\ne 0$, the singularity
becomes timelike. Although this seems similar to the
solution~\einmetric, there is a key difference. In \RN, if one evolves
the homogeneous initial data into the past, one finds a Cauchy
horizon. The solution can be extended beyond this horizon to an
asymptotically flat spacetime. This shows that the initial surface was
inside a black hole. For the solution~\einmetric, the initial data on
a constant $\tau < 0$ surface can be evolved infinitely far into the past
without reaching a Cauchy horizon.

We have not yet discussed the solution \einmetric\ near $r=0$.  In
this region, the $h\tau$ term is negligible. In terms of a proper
radial coordinate $\rho$ and a rescaled time coordinate $t$, the
metric (for fixed $\theta$ and $\varphi$) becomes
\eqn\rindler{ \ds = -\rho^2 \dt + {\rm d}\rho^2 \,.}
This is precisely the metric for Rindler space, {\it i.e.}~one of the
two regions outside the light cone of the origin of two dimensional
Minkowski spacetime, written in coordinates adapted to the boost
translation symmetry. It is now clear that $r=0$ (which is $\rho = 0$)
is a null surface consisting of two parts, corresponding to the future
and past Rindler horizon. This surface is also the location of a
curvature singularity since the area of the spheres vanishes there.
Thus we do not have an ideal counterexample to cosmic censorship.  All
(nonextendible) spacelike surfaces must hit $r=0$, and hence there is
no nonsingular initial data for this spacetime. The Penrose diagram
for the entire spacetime is shown in Fig.~2.

\ifig\bhfig{The Penrose diagram for the single mass solution. The
thick line at $r=0$ is the null singularity, and the thick
line at $r \tau = 2 M/h$ is the timelike singularity.}{\epsfbox{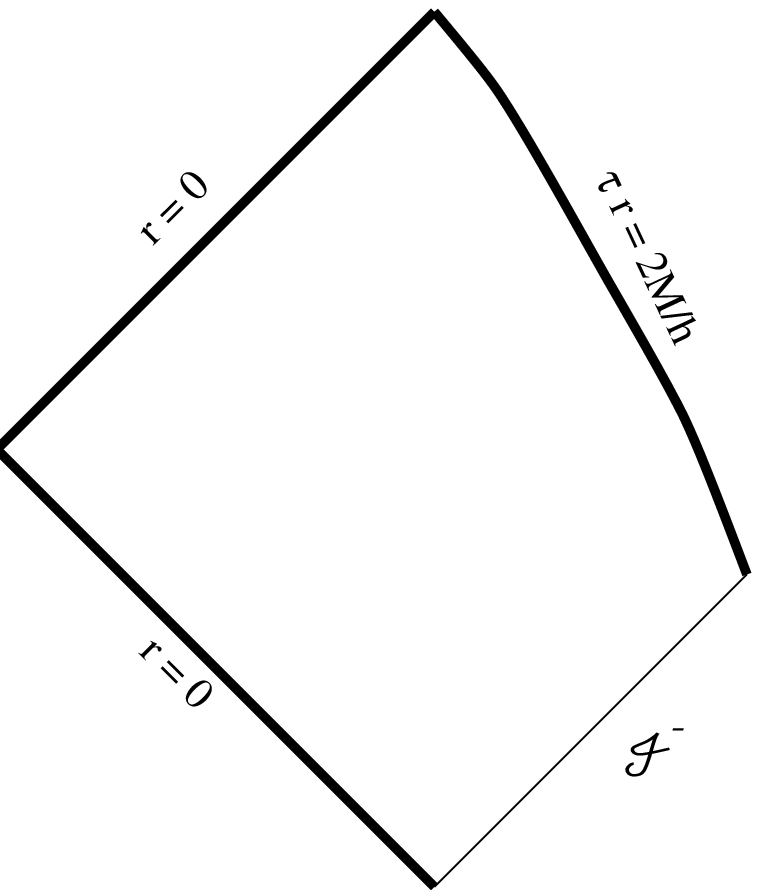}}

We do not believe this is a serious difficulty for the following
reason. As we have said, the solution near $r=0$ is similar to the one
with $h=0$ . The general charged dilatonic black hole with mass $M$,
charge $Q$, and $h=0$ can be expressed~\maeda,~\ghs
$$ \ds = - \(1-{2M\over \r}\) \dt + \(1-{2M \over \r}\)^{-1} {\rm d}\r^2
+ \r\(\r-{Q^2 \over
M} \){\rm d}\Omega^2 \,,$$
\eqn\einmet{ F_{\r t}={Q\over \r^2}\,,\qquad e^{2\phi} = 1-{Q^2\over M\r}\,.}
There is an event horizon at $\r= 2M$ and a spacelike singularity at
$\r= Q^2/M$.  In the extremal limit, $Q^2 = 2 M^2$, the event horizon
shrinks down to zero size and becomes singular. Since the causal
structure in the $\r,t$ plane is independent of the charge, it is
clear that this singularity is null and consists of two parts
corresponding to the future and past horizon. Setting $\r = r+2M$, the
extremal metric takes the form~\dmp\ with $\t U= 1 + 2 M/r$. So the
singularity at $\r=2M$ is directly analogous to the one we saw above
at $r=0$. This justifies the interpretation of \einmetric\ as an
extremal black hole in the collapsing universe.
But it is now clear that if we move slightly away from the
extremal limit and have $Q^2 < 2M^2$, then the surface $r=0$ will
become a nonsingular horizon shielding a spacelike singularity
inside.  Furthermore, the slight change in the charge should not
affect the solution asymptotically, so the naked singularity at large
$r$ should still be present. If so, one would then have nonsingular
initial conditions evolving into a naked singularity.  If we then
couple the theory to charged matter, we could presumably form a $Q^2 <
2M^2$ black hole by collapse. This would allow one to find violations
of cosmic censorship with a single asymptotic region.

Unfortunately, the usual techniques for adding charge to a vacuum
solution~\sen\ do not apply when a string cosmological constant is
present, and we have been unable to find a closed form expression for
the charged black hole in a universe with $h \ne 0 $ away from the
extremal limit. We have also been unable to solve the uncharged case
in the presence of $h$.  However, one can still construct nonsingular
initial data which evolve into a naked singularity by taking a surface
of constant $\tau <0$ in the solution \einmetric\ and restricting to
$r>r_0$ for some constant $r_0>0$. Since the timelike singularity
appears first at large $r$, it will still be obtained by evolving
this restricted initial data.  Of course, to obtain initial data which
are both nonsingular and complete, one needs to find the nonextremal
solution.  On the other hand, it appears likely that the naked
singularity at large $r$ does not depend on the exact nature of the
system near $r=0$.  If so, the formation of the timelike singularity
would be a universal feature for these black holes.

We now consider the string metric, which appears in~\saction. This
metric has rather different behavior. The single mass solution is
obtained by conformally rescaling the Einstein metric~\einmetric\ by
$e^{2 \phi} = 1/U$ to give
\eqn\stringgood{ \dhs = - { \dtau \over (\conf)^2} +
		\dr + \dOmega.}
Notice that the surfaces of constant $\tau$ are now completely flat. The
region $\tau < 0$ can be put into a somewhat simpler form by
introducing the coordinate $t$ measuring proper time at infinity,
$\tau = - e^{-h t}/h$. The metric is then
\eqn\stringbad{ \dhs = - { \dt \over (1 + 2 M e^{h t}/r)^2 } +
		\dr + \dOmega.}
This form of the metric was used in~\maki, but is somewhat misleading
since it does not cover the entire spacetime. If $M \ne 0 $, it is
possible to reach $t = \infty$ in finite affine parameter.  When
$M=0$, the solution is clearly just flat spacetime. The dilaton is
$e^{-2\phi_0} = -h\tau = e^{-ht}$ or $\phi_0(t) = h t /2$.
In other words, the $M=0$ limit is just the familiar linear dilaton
solution~\mye\aben. Even when $M \ne 0$, \stringbad\ clearly approaches the
linear dilaton solution asymptotically. In fact, the contribution from
the mass goes to zero exponentially as $t \a -\infty$. So the
spacetime is asymptotically flat in the usual sense at past null
infinity and has a complete $\Im^-$.

Even though the spacetime is asymptotically flat, the standard methods
of defining the total mass do not apply since the asymptotic dilaton
is time dependent. The charge is defined as usual to be $Q= {1\over
4\pi} \oint e^{-2\phi} {}^* F$ where $*$ denotes the dual, and the
integral is over a large sphere at infinity. From the solution
for the vector potential~\dkt, one can verify that $Q =\sqrt 2 M$
where $M$ is the parameter in the solution.

Just like the Einstein metric \einmetric, the curvature of the string
metric~\stringgood\ diverges at $r=0$ and $r \tau = 2 M/h$. However,
the causal structure is quite different.  Timelike and null geodesics
never actually reach the timelike singularity at $r \tau = 2 M/h$.
Roughly speaking, this is because the singularity occurs at $U=0$, and
the string metric is obtained by multiplying the Einstein metric by
$U^{-1}$.  Thus the conformal factor becomes very large in this region
and essentially pushes the singularity off to infinity.

We verify this by again considering the geodesics.  As before, let
$\lambda$ denote an affine parameter, and set $\dot{\theta} =
\dot{\phi} = 0$. Then geodesics must satisfy
\eqn\stringgeoda{ - \kappa = - { \dot{\tau}^2 \over (\conf)^2} + \dot{r}^2,}
where $\kappa = 1,0,-1$ for timelike, null, and spacelike geodesics.
The geodesic equation for $r$ now becomes
\eqn\stringgeodb{ \ddot{r} + { 2 M (\dot{r}^2 + \kappa) \over r^2 (\conf)}
		= 0\,.}
First consider null geodesics with $\kappa = 0$. Unlike the simple
solution available for the Einstein geodesics~\eingeod, it is not
possible to solve~\stringgeodb\ in closed form. We can determine the
behavior of null geodesics very near the singularity though. As an
outgoing null geodesic gets very near the timelike singularity,
\stringgeoda\ forces $\dot{\tau}$ to go to zero. Thus $\tau$
approaches a constant near the singularity.  In terms of a new
coordinate $z = r - {2 M \over h \tau}$ which vanishes at the
singularity and is negative near it, the geodesic
equation~\stringgeodb\ approaches
\eqn\zgeod{ \ddot{z} - {\dot{z}^2 \over z} = 0\,,}
where we have dropped terms that go to zero at the singularity.  This
equation can easily be solved, to get $z = - e^{-\lambda}$.  Thus null
geodesics never reach the timelike singularity! Instead, they approach
it exponentially slowly. Both $\dot{\tau} \a 0$ and $\dot{r} \a 0$ for
null geodesics near the singularity.

Timelike geodesics are deflected before they reach the singularity.
Consider~\stringgeodb\ with $\kappa = 1$. Since $\conf > 0$, the
$\kappa$ piece contributes an extra negative acceleration compared to
a null geodesic. Thus instead of $\dot{r} \a 0$, $\dot{r}$ becomes
negative before reaching the singularity, and the timelike geodesics
move away from the timelike singularity. Thus neither timelike nor
null geodesics ever reach the timelike singularity, and the spacetime
is causally geodesically complete (for outgoing geodesics).

For spacelike geodesics, it follows from~\stringgeoda\ that
$\dot{r} > 1$ everywhere.  Since the singularity is at finite $r$, this
implies that all outgoing spacelike geodesics reach the singularity in
finite affine parameter, and the spacetime is spacelike geodesically
incomplete.

Since the theory \saction\ expressed in terms of the string metric is
related to the one expressed in terms of the Einstein metric \eaction\
by a field redefinition, they should represent the same physics. Yet,
in one metric the $r\tau = 2M/h$ singularity is at infinity, and in
the other metric it is not. So does this solution have a physical
naked singularity? If additional matter is present and minimally
coupled to the Einstein metric, the answer is clearly yes. This matter
can propagate from the singularity in finite time.  Even without
additional matter, one can argue that gravitational waves will always
propagate with respect to the Einstein metric, and hence one could see
the effects of the singularity this way.

\newsec{Generalizations of the single mass results}

The multi-mass solution in the Einstein metric, eqs.~\udef\ and~\dkt,
has features similar to the single mass solution discussed in the
previous section.  For $\tau < 0 $ the solution describes a number of
extremal black holes in a contracting universe.  Since $U$ is a
decreasing function of $\tau$, the proper distance between the black
holes,
\eqn\pdist{ l = \int_{r_i}^{r_j} \sqrt{U} dl\,,}
is also a decreasing function of $\tau$ (in the string metric, the
proper distance between the masses remains constant).  Thus the black
holes are approaching each other.  For large $r$ the solution reduces
to the single mass solution with $M= \sum M_i$. So as before, when
$\tau = 0$ a timelike singularity appears at $r = \infty$ and moves in
toward smaller $r$. This singularity is located at $U=0$. So as $\tau$
increases,
it eventually splits, and surrounds
each of the masses separately.

The multi-mass solution can also be extended to spacetimes with
dimension $d > 4$~\maki. The solution in the Einstein metric is again
given by~\dkt, but with a generalized $U$ given by
\eqn\uddef{  U = -h\tau+\sum_i {2 M_i\over |\vec{r}-\vec{r}_i|^{d-3}}\,.}
The higher dimension solution is qualitatively similar to the four
dimension case. The geodesic equation is again~\nullddr, so the
spacetime again has a naked timelike singularity. In the string metric,
null geodesics near the singularity still approach~\zgeod, so even in
higher dimensions, null geodesics in the string metric do not reach
the singularity.

We have considered only masses with electric charge. It turns out that
adding magnetically charged extremal black holes to the collapsing
universe~\flateina\ does not create naked singularities.  As mentioned
earlier, the magnetically charged solution cannot be obtained by
applying a simple duality transformation to~\einmetric. However, it
can be obtained by recalling that the string metric for the $h=0$
solution is
\eqn\stmag{ \dhs = -\dht + \t U^2 \drv\,,}
where $\t U$ is given by \upm, and the dilaton is $e^{2\phi} = \t U$.
In other words it is simply the product of time and a three
dimensional spatial solution. Since the action \saction\ comes from
the condition of two dimensional conformal invariance, and the product
of conformal field theories is again a conformal field theory, it
follows that a solution with $h\ne 0$ can be obtained by taking the
product of the same spatial solution and a linear dilaton solution in
$\t t$. Thus the $h \ne 0$ solution has the same string metric \stmag,
and the dilaton given by $e^{2\phi} = e^{h \t t} \t U$ \maki.  The
string metric remains static and nonsingular. Rescaling back to the
Einstein metric and introducing the new time coordinate $ t = -{2
\over h} e^{-h \t t/2}$ yields
\eqn\eimag{ \ds = -{\dt \over \t U} + {1\over 4} h^2 t^2 \t U \drv\,.}
For large $r$, this metric resembles the electrically charged solution
with $\tau < 0$. However, now the null singularity at $t=0$ remains,
and is not converted into a timelike singularity.

\newsec{Conclusions}

We have discussed a family of solutions to the
Einstein-Maxwell-dilaton theory~\eaction\ which can be viewed as the
dilatonic analog of the recently discovered Kastor-Traschen solution.
The solution describes an arbitrary number of extremal electrically
charged black holes in a contracting universe. In the absence of the
black holes, the universe collapses to a null singularity. But when a
single extremal black hole is added, the null singularity turns into a
timelike one which is naked. This suggests that cosmic censorship may
fail for this theory. To further explore whether this is the case, it
would be useful to find the solution describing nonextremal charged
black holes in the contracting universe. One could then verify that
the naked singularity which arises at large $r$ is independent of the
details of the solution at small $r$. It is also important to study the
stability of these solutions. This may depend on the boundary conditions
imposed at infinity. Clearly, perturbations of compact support cannot
affect the formation of the naked singularity.  Physically reasonable
boundary conditions should allow one to add black holes of arbitrary
mass and charge to the spacetime.

Since the theory we are considering is the low energy limit of string
theory (with a central charge), one is led to ask whether cosmic
censorship is violated in string theory. Unfortunately, we can say
very little at this time.  In the context of string theory, the region
of the solution near $U=0$ cannot be trusted. Since the
curvature is becoming large, higher order $\alpha'$ corrections will
be important.
 Since string theory is in principle a
complete quantum theory of gravity, a breakdown of cosmic
censorship does not imply a breakdown of the theory, but simply that
quantum corrections must be included. Indeed, in our solutions,
the string coupling $g = e^\phi$ is becoming
large near the naked singularity and string loop corrections will be
important.

Finally, it is worth remarking that the dilaton may not be necessary
to create the naked singularities we have found here. If one considers
Einstein's equation with a perfect fluid source having equation of
state $p=-\rho/3$, one has the solution \flateina\ with its null
singularity. If we add a black hole to this solution, does the
singularity become timelike?

\bigskip\centerline{Acknowledgements}

\bigskip
It is a pleasure to thank D.~Garfinkle for discussions.  This work was
supported in part by NSF Grants PHY-8904035 and PHY-9008502, and by
DOE grant DE-AC02-76ER03075.

\listrefs

\bye